# Resistance noise at the metal–insulator transition in thermochromic VO$_2$ films

Zareh Topalian [1], Shu-Yi Li [1], Gunnar A. Niklasson [1], Claes G. Granqvist [1], and Laszlo B. Kish [1,2,a]

[1] *Department of Engineering Sciences, The Ångström Laboratory, Uppsala University, P. O. Box 534, SE-75121 Uppsala, Sweden*

[2] *Department of Electrical and Computer Engineering, Texas A&M University, College Station, TX 77843-3128, USA*

Thermochromic VO$_2$ films were prepared by reactive DC magnetron sputtering onto heated sapphire substrates and were used to make 100-nm-thick samples that were 10 $\mu$m wide and 100 $\mu$m long. The resistance of these samples changed by a factor ~2000 in the 50 < $T_s$ < 70 ºC range of temperature $T_s$ around the "critical" temperature $T_c$ between a low-temperature semiconducting phase and a high-temperature metallic-like phase of VO$_2$. Power density spectra $S(f)$ were extracted for resistance noise around $T_c$ and demonstrated unambiguous $1/f$ behavior. Data on $S(10\ \mathrm{Hz})/R_s^2$ scaled as $R_s^x$, where $R_s$ is sample resistance; the noise exponent $x$ was –2.6 for $T_s < T_c$ and +2.6 for $T_s > T_c$. These exponents can be reconciled with the Pennetta–Trefán–Reggiani theory [C. Pennetta, G. Trefán, and L. Reggiani, Phys. Rev. Lett. **85**, 5238 (2000)] for lattice percolation with switching disorder ensuing from random defect generation and healing in steady state. Our work hence highlights the dynamic features of the percolating semiconducting and metallic-like regions around $T_c$ in thermochromic VO$_2$ films.

………………………
[a] Corresponding author. Electronic mail: laszlo.kish@ece.tamu.edu.

# I. INTRODUCTION

This paper shows that $VO_2$ films display strong electrical resistance noise around their "critical" temperature $T_c$ for thermochromic switching, and that this phenomenon can be reconciled with scaling concepts.

The thermochromism of $VO_2$ was discovered many years ago.[1] This phenomenon is associated with a first-order metal–insulator transition (MIT) at $T_c \approx 68$ ºC, and $VO_2$ is capable of switching between a low-temperature (monoclinic, M1) semiconducting state and a high-temperature (rutile, R) metallic-like state. Thin films can display highly reversible switching and are of intense current interest for numerous applications such as glazings for energy-efficient buildings[2–5] and a large number of (opto)electronic, bolometric, and sensing devices.

In single crystals of $VO_2$, the MIT is characterized by sharp real-space phase boundaries between the R and M1 structures and by the fact that these boundaries can propagate along the crystallographic *c*-axes,[6] *i.e.*, the transition is non-percolating in nature. Thin films of $VO_2$ are normally distinctly different, however, and the MIT is gradual with metallic-like regions growing in extent as the sample temperature $T_s$ approaches $T_c$ from below and with semiconducting regions disappearing as $T_s$ becomes increasingly larger than $T_c$. This behavior has been convincingly demonstrated via a combination of scanning near-field infrared microscopy, atomic force microscopy, and x-ray diffraction,[7] and analogous results have been inferred from other data.[8–13] The percolative character of the MIT in $VO_2$ films has been emphasized several times.[9,14–17] It should be noted that the properties of $VO_2$-based materials around the MIT are not only of theoretical interest but influence, for example, the energy-savings potential of thermochromic glazings.[18]

Percolation enhances macroscopic resistance fluctuations, as is well known,[19] and such fluctuations have been investigated in some prior studies[20–23] especially with regard to bolometer performance. Our present investigation goes well beyond earlier work, as far as we know, and reports on a detailed study of resistance noise around the MIT in $VO_2$ films. Sample preparation is discussed in Sec. II. The experimental set-up for measuring noise is described in Sec. III, where noise data are given as well. The theoretical background for analyzing these data is given in Sec. IV, where we also introduce several earlier theories for the noise. Section V contains a discussion of the data and shows that the Pennetta–Trefán–Reggiani theory[24] can be reconciled with the experimental results.

## II. SAMPLE PREPARATION AND CHARACTERIZATION

### A. Sputter deposition

Thin films of $VO_2$ were prepared by reactive DC magnetron sputtering in a deposition system based on a Balzers UTT 400 unit. The chamber was evacuated to a base pressure of $6.3 \times 10^{-7}$ mbar by turbo molecular pumping, and 80 ml/min of Ar and 5 ml/min of $O_2$ (both 99.997% pure) were then introduced through a mass-flow-controlled gas inlet; the total pressure was maintained at $1.2 \times 10^{-2}$ mbar during the deposition. Sputtering was performed from a 5-cm-diameter target of vanadium (99.5% pure) at a power of 172 W onto sapphire substrates, 25 mm in diameter and 0.5 mm thick, heated to 450 °C. The film thickness was 100 nm, as determined by use of a Bruker DektakXT profilometer, and the film growth rate was found to be ~0.06 nm/s. Further details on the deposition technology are given elsewhere.[25,26]

### B. Sample configuration

Resistance noise measurements on materials with modest disorder require small sample volumes in order to avoid excessive heating. We therefore made structures comprising a narrow $VO_2$ micro-bridge by photolithography combined with reactive ion etching (RIE). A four-contact arrangement was used to reduce contact noise. Lithography was employed to transfer the geometric shape defining the sample, printed on a mask and shown in Fig. 1(a), to the surface of the substrate. The remaining uncovered part of the $VO_2$ film was removed by RIE using a system based on a Vision 320 unit (Advanced Vacuum, Sweden). For the etching process, a gas flow with 5 ml/min of $O_2$ and 45 ml/min of $CHF_3$ was introduced into the RIE unit and the total pressure was kept at 0.027 mbar. An RF power of 400 W at 13.56 MHz was applied to generate plasma during 40 minutes of etching time. Figures 1(b) and 1(c) are images of the resulting $VO_2$ structure, taken with a digital camera from a close distance and with an Olympus BX60 microscope at 100× magnification, respectively.

Figure 2 shows scanning electron microscopy (SEM) images on the entire $VO_2$ micro-bridge and of small portions at the center and edge of this structure. Data were taken with a Zeiss LEO 1550 instrument equipped with an energy dispersive x-ray spectroscopy (EDS) system for elemental analysis. The micro-bridge is found to be 10 $\mu$m wide and 100 $\mu$m long. Clearly the $VO_2$ structure is well defined, uniform, and characterized by granular features on the scale of ~50 nm. EDS verified that the etched regions did not show any traces of residual vanadium.

### C. Temperature-dependent resistance

The thermochromic properties of the $VO_2$ micro-bridge were verified by measurements of electrical resistance $R_s$ during heating and cooling in the 20 < $T_s$ < 80 ºC interval. The temperature recording had to rest at least for 10 minutes at each point in order to stabilize the $T_s$ and resistance readings. Figure 3 shows that $R_s$ changes by a factor ~2000 in the range between ~50 and ~70 ºC and that the transition displays thermal hysteresis amounting to ~7 ºC.

## III. EXPERIMENTAL RESISTANCE NOISE

### A. Measurement technique

Measurements of spontaneous resistivity fluctuations (noise) are technically challenging for $VO_2$ as a consequence of its high temperature coefficient. We note, in passing, that this property constitutes the foundation for applications of $VO_2$ in (micro)bolometers[22,27–31] and amplifiers utilizing stochastic resonance.[32,33] The most temperature-dependent component in the measurement system is the $VO_2$ sample itself, and its conductivity is prone to display unwanted temperature fluctuations caused by the temperature control unit. In order to avoid this artifact in the 1–200 Hz frequency range of primary interest we used a sample holder, comprised of a copper disc positioned in vacuum, with large thermal capacitance and weak coupling to the rest of the system; this resulted in a thermal relaxation time constant greater than 1000 s. The temperature control system was specially designed in order to meet the stringent demands of our resistance noise measurements.[34–38]

Resistance noise in the temperature-stabilized $VO_2$ sample was measured while running a DC current through it by use of a 12 V car battery and low-noise resistors with proper resistance values. Background noise, recorded at zero DC current drive, was subtracted. The ensuing mean-square value of the voltage fluctuations within a narrow pass band around the frequency $f$ and normalized by the band width—*i.e.*, the noise spectrum $S_R(f,T_s)$—was then obtained in the standard way by linear circuit theory.[39] We checked that $S_R(f,T_s)$ did not depend on the applied DC current, thus confirming that the measured excess noise was indeed due to resistance fluctuations. Furthermore, at fixed DC current, the measured $S_R(f,T_s)$ was independent of the value of the current generator's resistance, which shows that effects of contact noise were negligible.

## B. Data

Figure 4 reports a measured noise spectrum $S_R(f,T_s = 48.8\ °C)$ for the VO$_2$ sample depicted in Fig. 2, recorded for decreasing temperature. The data show a frequency dependence that could be fitted to a $1/f^\gamma$-behavior with $\gamma \approx 1.1$ at the given temperature; this is the temperature at which the absolute value of the sample's temperature coefficient of resistance has its maximum, which is about $10^5$ times greater than at room temperature. The data obviously deviate strongly from a $1/f^2$ behavior, which would have been expected from artefacts stemming from the temperature control unit.[40]

We then measured $S_R(f = 10\ Hz, T_s)/R_s^2$, *i.e.*, a resistance noise spectrum at 10 Hz and normalized by the square of the sample resistance, as $T_s$ was varied around $T_c$. Figure 5 shows corresponding data on $S_R(f = 10\ Hz, R_s)/R_s^2$ and demonstrates that power-law scaling *versus* $R_s$ prevails over several orders of magnitude in $R_s$ for the low-temperature (high-resistance) end as well as the high-temperature (low-resistance) end of the data. These measurements were taken during heating and cooling. The exponents in the noise data are found to be approximately –2.6 at $T_s < T_c$ and +2.6 at $T_s > T_c$, and a transition zone is apparent around $T_c$.

## IV. THEORETICAL MODELS FOR RESISTANCE NOISE

### A. Background

Percolation around the MIT in VO$_2$ films is characterized by the mean size of interconnected semiconducting and metallic-like regions and their structural features. From a percolation point view, it is appropriate to call the transition a good-conductor–bad-conductor (GCBC) transition wherein the good-conducting phase is increasingly approaching the percolation threshold in the low-temperature phase during heating, whereas the bad-conducting phase is increasingly approaching the percolating threshold in the high-temperature phase during cooling. In both cases, the characteristic size of regions in the percolating phase is referred to as the "correlation length" $l_c$ which, in an infinite system with self-similar structure, scales according to

$$l_c \propto (p - p_c)^{-r} \qquad (1)$$

where $p$ is the volume fraction ($0 \leq p \leq 1$) of the good-conducting regions, $p_c$ is the percolation threshold ($0 < p_c \leq 1$), and the percolation exponent $r$ often is universal. The correlation length diverges at $p_c$ and, close to this threshold, macroscopic electrical transport properties display power-law scaling with $l_c$. Earlier work[9,14–16] showed the existence of percolation but was unable to give finer details about this phenomenon.

Equation (1) also implies that transport properties scale with $(p - p_c)$ and hence have power-law scaling with each other. Such quantities are, for example, resistance and normalized resistance fluctuations $\langle \Delta R^2(t, T_s) \rangle / R_s^2(T_s)$, where $\Delta R(t, T_s)$ is resistance noise at $T_s$ and time is denoted $t$. Thus one way to probe the existence and nature of percolation in random conductance structures is to measure resistance noise and check whether there is power-law scaling between the resistance and the strength of its fluctuations while temperature is varied.

Energy conservation, as encapsulated in the Cohn–Tellegen theorem,[41–44] relates the macroscopic resistance and resistance fluctuations to microscopic quantities. If the microscopic mean-square resistance fluctuations are small in the sub-volumes (*i.e.*, regions in VO$_2$), the contributions of these resistances are added linearly in the total resistance with weighting factors proportional to the local current through the sub-volume. Thus

$$R_s = \frac{\sum_k r_k I_k^2}{I^2}, \qquad (2)$$

where $I = \left( \sum_k I_k \right)$ is the mean DC current through the sample, $r_k$ is the resistance of the $k$-th sub-volume, and $I_k$ is the mean DC current flowing through it. Due to the uncorrelated nature of the resistance fluctuations within these sub-volumes, the power density spectrum (noise spectrum) $S_R(f)$ of $\Delta R(t)$ is[44]

$$S_R(f) = \frac{\sum_k s_k(f) I_k^4}{I^4}, \qquad (3)$$

where $s_k(f)$ is the noise spectrum of the resistance fluctuations of $r_k$. Hence the noise probes the fourth moment of the current density distribution while the resistance probes the second moment.[19] The normalized resistance noise spectrum of the total resistance is then given by

$$\frac{S_R(f)}{R_s^2} = \frac{\sum_k s_k(f) I_k^4}{\left(\sum_k r_k I_k^2\right)^2} \ . \qquad (4)$$

For classical lattice-percolation with a periodic lattice, fixed $s_k = s \ (k = 1, ...)$ values, and sufficiently close to the percolation threshold, there is scaling according to[19,44–46]

$$\frac{S_R(f)}{R_s^2} \propto R_s^x \ , \qquad (5)$$

where the noise scaling exponent $x$ is universal and dependent on the dimensionality of the percolation.

**B.  Scaling exponents for resistance noise**

In *lattice percolation* and noise generation due to uncorrelated resistance noise of non-overlapping sub-volumes, $x$ is about +1 for percolation of the bad-conductor phase (often called "insulator" in the literature) through the good-conductor phase, whereas $x$ is about −1 for percolation of the good-conductor phase (often called "superconductor") through the bad-conductor phase, but the exact values of $x$ are dimensionally dependent.[44,46,47] It should be noted that these theoretical models attribute the noise to the non-percolating phase, and the percolating phase is taken to be noise-free irrespectively of its being "insulator" or "superconductor".

Noise in lattice percolation that is not generated by resistance fluctuations in the sub-volumes but instead by *random switching events* in the value of $p$ with constant or weakly changing variance during the transition is another possibility[46] and is referred to as "$p$-noise". In this case, a GCBC transition with percolating good-conductor has universal exponents $x$ that are −2.7 in three dimensions (3D), −1.5 in 2D and −2 in 1D, whereas a GCBC transition with percolating bad-conductor has $x$ values of +1 in 3D and +1.5 in 2D.

The Pennetta–Trefán–Reggiani (PTR) lattice percolation noise model[24] operates with *randomly switching defects* in the lattice. A given lattice site has a *failure probability* to malfunction and become a defect and a *healing probability* to for the defect to recover its original lattice resistance. The generated noise is switching-based, just as in the *p*-noise model, but the variance of the fluctuations in *p* is not constant. When a steady state is reached, the difference of the two probabilities yields the mean value of *p* in a non-trivial way. According to computer simulations in 2D, the resulting exponent is about 2.6 in the case when the defects are insulating lattice elements, which is the only situation that so far has been studied.

Exponents that are non-universal and much larger than those mentioned above have been found in *continuum percolation* described by special models, such as in the "Swiss cheese" model with overlapping insulating spherical or circular voids.[19,45] For example, Garfunkel and Weissman[19] found *x*-values between 2 and 4 for different versions of the "Swiss cheese" model. Experimental data on sandblasted metal films, which were interpreted within such a model in 2D, could be described by $3.4 < x < 6.1$.[19]

## V. DISCUSSION OF RESISTANCE NOISE

### A. Disorder assessment in uniform phases

We first discuss the strength of the $1/f$-type noise in VO$_2$, shown in Fig. 4, in the single-conductance phase limits, *i.e.*, in the pure bad-conductor (low-temperature) phase and in the pure good-conductor (high-temperature) phase. For a sample with homogeneous cross-section and current density, and when the noise is due to independent elementary fluctuating sub-volumes homogeneously distributed in the sample volume, the normalized noise spectrum $S_R(f,T_s)/R_s^2(T_s)$ *versus* resistance can be given as[48,49]

$$\frac{S_R(f,T_s)}{R_s^2(T_s)} = \frac{\alpha_m(T_s)}{N_m f^\gamma} , \qquad (6)$$

where $N_m$ is the number of "molecules" in the sample and $\alpha_m(T_s)$ is the parameter characterizing the "molecular" strength of bulk $1/f$ noise in the material.

The well-known Hooge formula[50,51] is often used under similar conditions and states that

$$\frac{S_R(f,T_s)}{R_s^2(T_s)} = \frac{\alpha_H(T_s)}{N_e f^\gamma} , \qquad (7)$$

where $N_e$ is the number of conducting electrons or other charge carriers in the sample and $\alpha_H$ is the "Hooge parameter". This parameter often serves as a useful measure for comparing the strength of $1/f$ noise in different materials, even though the general theory underlying Eq. (7) has been questioned.[52,53] Equations (6) and (7) imply that the two measures on the strength of $1/f$ noise at homogeneous current density can be expressed from the measured quantities by

$$\alpha_m(T_s) = N_m f^\gamma \frac{S_R(f, T_s)}{R_s^2(T_s)} \tag{8}$$

and

$$\alpha_H(T_s) = N_e f^\gamma \frac{S_R(f, T_s)}{R_s^2(T_s)}. \tag{9}$$

We can now make use of data on the electron concentration $n_e$ in semiconducting (subscript *es*) and metallic-like (subscript *em*) VO$_2$—which are $n_{es} \approx 10^{18} - 10^{19}$ cm$^{-3}$ and $n_{em} \approx 3.3 \times 10^{22}$ cm$^{-3}$, respectively[54]—and the sample volume which is $10^{-10}$ cm$^{-3}$. This leads to $N_{es} \approx 10^8$ for the lower value of $n_{es}$ and $N_{em} \approx 3 \times 10^{12}$. Equations (8) and (9) then yield $\alpha_{ms} \approx 3.8$ and $\alpha_{mm} \approx 3.8 \times 10^{-2}$, and the corresponding Hooge parameters are $\alpha_{Hs} \approx 1.3 \times 10^{-4}$ and $\alpha_{Hm} \approx 3.8 \times 10^{-2}$. All of these values characterize a film with medium-strong disorder.[48,49,51,55]

## B. Interpretation of noise exponents

We first note that the experimental data interpreted within the "Swiss cheese" model in 2D and the *p*-noise model in 3D, introduced in Sec. IV B, both yield exponents $x$ with values that are similar to those observed in our measurements. Hence the exponents pertinent to the "Swiss cheese" model[19,45] might explain the *positive* exponent in the high-temperature regime where the bad-conductor percolates, but it is difficult to imagine that this model would be applicable for the high-resistance regime with the *negative* exponent where the good-conductor percolates, even if duality would theoretically allow for this possibility in 2D.[47] The formation of a hypothetical "Swiss-cheese"-like structure in the latter case would imply that good-conductor regions in VO$_2$ could somehow be overlapping, which appears unphysical or at least highly unlikely.

Similarly, the *p*-noise model[46] for superconductor-percolation, which yields $x \approx -2.7$ in 3D, might explain the negative exponent in our experiment, but this model fails to produce the desired large positive exponent for insulator-percolation, which is instead predicted to be characterized by $x \approx 1$.

The only model that so far seems able to account for the observed large noise exponents in both the high-temperature and low-temperature regimes is the PTR model[24] for switching disorder with random defect generation and healing under steady-state conditions. This model was deduced for a bad-conductor phase that percolates trough a good-conductor phase, and 2D simulations yielded that $x \approx 2.6$. PTR did not study the case of good-conductor percolation in a bad-conductor phase, but the duality of 2D percolation networks predicts[47] that $x \approx -2.6$ for the latter case.

Finally we note that dimensionality cross-over is possible in percolating systems.[56–58] Specifically 2D behavior is valid sufficiently close to $p_c$, where the correlation length is larger than the film thickness. No effect of dimensionality cross-over could be documented in our experimental data, however, which may be consistent with the sizes of the $VO_2$ regions of different types that were detected by scanning near-field infrared microscopy, atomic force microscopy and x-ray diffraction,[7] and referred to in the Introduction.

## VI. CONCLUSION

We prepared 100-nm-thick thermochromic $VO_2$ films by reactive DC magnetron sputtering onto sapphire substrates heated to ~450 ºC and employed them in 10-$\mu$m-wide and 100-$\mu$m long micro-bridge samples. Their thermochromism was documented by measurements of temperature-dependent resistance, which showed a change by a factor ~2000 in the 50–70 ºC range. Thus thermochromism is well developed and, in fact, a vast number of studies reported in the literature have been made on $VO_2$ samples with less distinct and wider transitions between the low-temperature and high-temperature states.

Power density spectra were extracted for resistance noise around $T_c$ and showed clear evidence for $1/f$ behavior. Data on the noise power at 10 Hz, divided by $R_s^2$, scaled as $R_s^x$ where the $x$ was –2.6 for $T_s < T_c$ and +2.6 for $T_s > T_c$. These exponents can be reconciled with the Pennetta–Trefán–Reggiani theory[24] for lattice percolation with switching disorder due to random generation and healing of defects under steady-state conditions. Generation and healing of these defects correspond to time-dependent transitions between semiconducting and metallic-like parts in $VO_2$ films. Our work therefore sheds light on *dynamic* features of the percolating semiconducting and metallic-like regions around the metal–insulator transition in thermochromic $VO_2$ films.

## ACKNOWLEDGMENTS

We are grateful to Per Nordblad and Peter Svedlindh for their assistance with the ultra-low noise temperature control system. Hans-Olof Blom is thanked for guidance in the application of the etching process. Financial support was obtained from the European Research Council under the European Community's Seventh Framework Program (FP7/2007–2013)/ERC Grant Agreement No. 267234 (GRINDOOR). LK's last (fall, 2014) visit to Sweden was partially founded by Texas A&M University.

[38]The sample holder had an attached thermometer which was constructed as a bifilarly wound coil of 0.08-mm-diameter Cu wire in order to reduce thermal noise and electromagnetic interference. This thermometer employed four-point resistance measurement utilizing the temperature dependence of the resistivity of the Cu wire. The current contacts of the thermometer were driven by an AC current generator formed by a high-precision large resistor and the oscillator output of a lock-in amplifier (Princeton Applied Research, type PAR 5210). The current generator's resistor was located in a thermally insulated and electrically screened oil bath at room temperature; the thermal time constant of this oil bath was similar to that of the sample holder. One side of the temperature-control bridge was the secondary coil of a precision transformer; the primary coil of this transformer was driven by the voltage electrodes of the thermometer. The other side of the temperature-control bridge was the reference voltage given by the secondary coil of a Decatron programmable transformer system. The primary coil of the Decatron unit was driven by the oscillator output of the same lock-in amplifier. The purpose of the lock-in amplifier was to enlarge the difference between the output voltages of the two secondary coils in the bridge. The lock-in oscillator was operated at 473 Hz in order to avoid 50 Hz harmonics, and notch filters were used for both 50 and 100 Hz. The time constant of the lock-in amplifier was set to 1 s with 12

dB/octave slope, and hence the effective bandwidth of the temperature measurement system was about 0.15 Hz at 473 Hz with about 72 dB/Hz theoretical cut-off efficiency. The error voltage, *i.e.*, the output voltage of the lock-in amplifier, drove a custom-designed and optimized PID controller whose D channel was disabled in order to improve the noise properties of the control. This PID unit was connected to a DC heater amplifier, which heated the sample holder by using a bifilar coil of manganine wire.

**FIGURE CAPTIONS**

FIG 1. Panel (a) is a schematic illustration of a 100-nm-thick VO$_2$ sample with a micro-bridge in the middle and with contact pads for four-point electrical measurements. The dimensions are $L_1$ = 5 mm, $L_2$ = 100 $\mu$m, $L_3$ = 6 mm, $d$ = 1.5 mm, $W_1$ = 1 mm and $W_2$ = 10 $\mu$m. Panels (b) and (c) are photos of the same structure and of the VO$_2$ micro-bridge in the encircled region, respectively. Copper wires were attached to the contact pads by silver glue, and resistance and noise measurements were performed on the sample. The left-hand contacts were fed via a DC current generator and the right-hand contacts were used for DC voltage and noise voltage measurements, as discussed in detail in the main text.

FIG. 2. SEM micrographs illustrating the VO$_2$ micro-bridge, which is also shown in Fig. 1(c). Panels (a), (b), and (c) depict the whole structure, a centrally positioned portion of it and an edge part, respectively.

FIG. 3. Temperature-dependent resistance during heating and cooling of a VO$_2$ film comprising the micro-bridge in a sample according to Figs. 1 and 2. Symbols represent measured data and connecting lines were drawn for convenience.

FIG. 4. Log-log display of a resistance noise spectrum, as explained in the main text, measured on a VO$_2$ film in the sample configuration shown in Figs. 1 and 2. The slope was –1.1, which clearly deviates strongly from a slope of –2 shown by a separate line. Some interference from the power line is noticeable around 50 Hz.

FIG. 5. Log-log display of normalized resistance fluctuations at 10 Hz, as explained in the main text, *versus* sample resistance $R_s$. Sample temperature is the hidden control parameter. The normalized noise at fixed resistance value was the same during cooling and heating. Scaling characteristics are indicated by lines with slopes relevant for the Pennetta–Trefán–Reggiani model, as elaborated later.

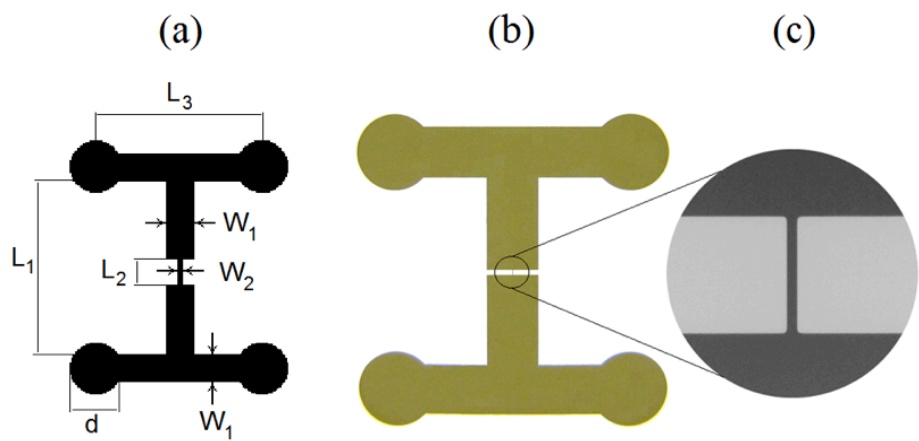

Fig. 1.

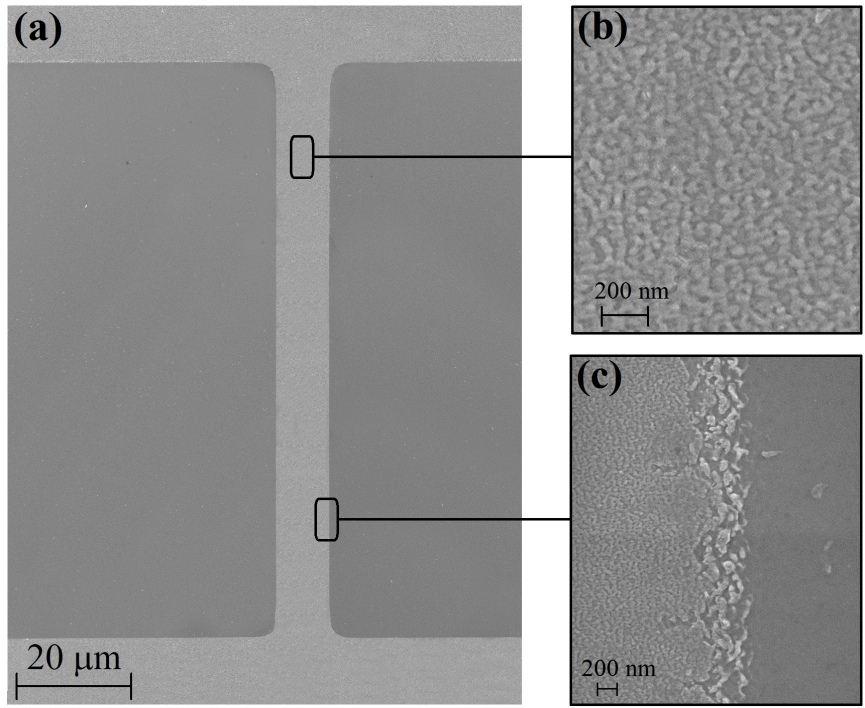

Fig. 2

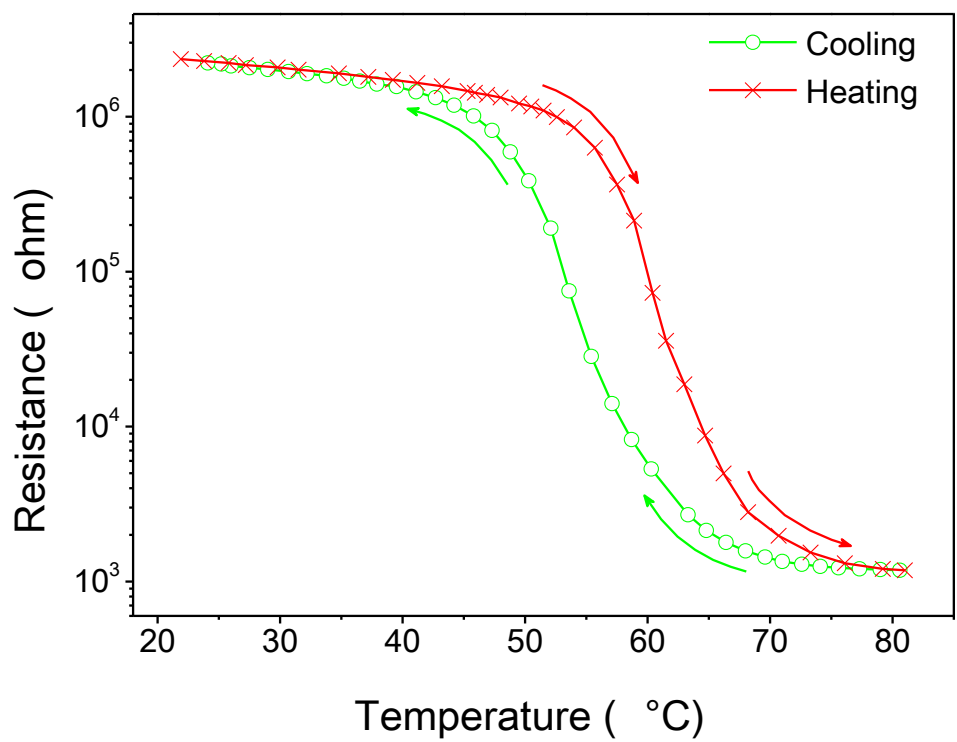

Fig. 3.

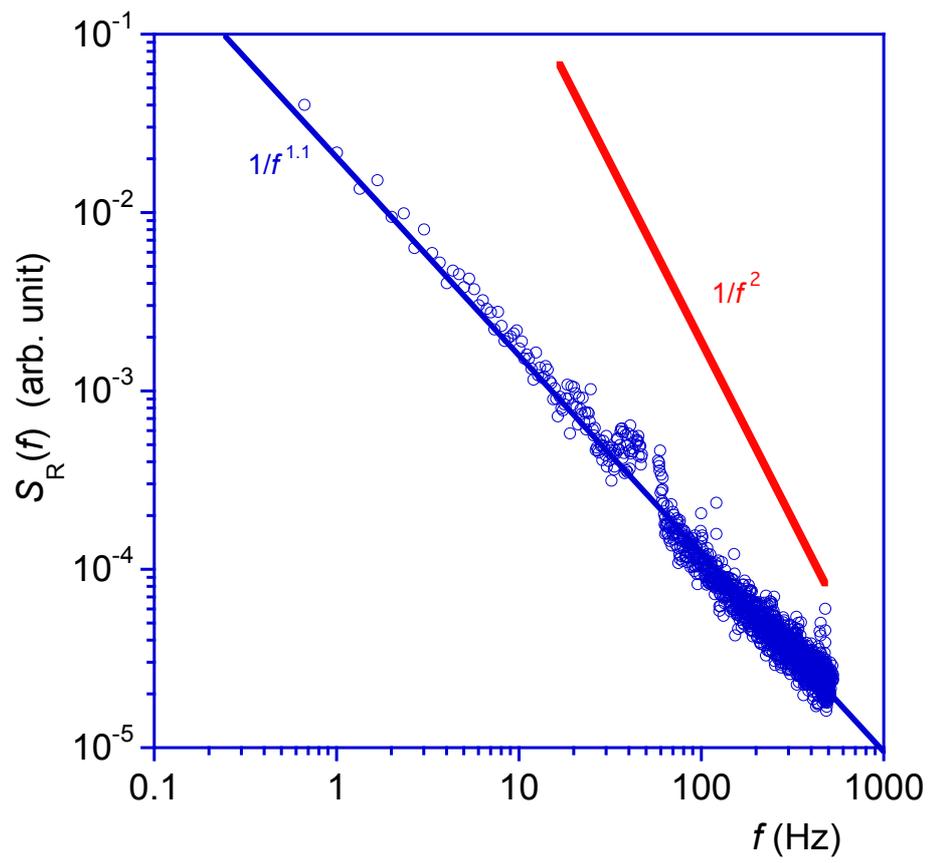

Fig. 4.

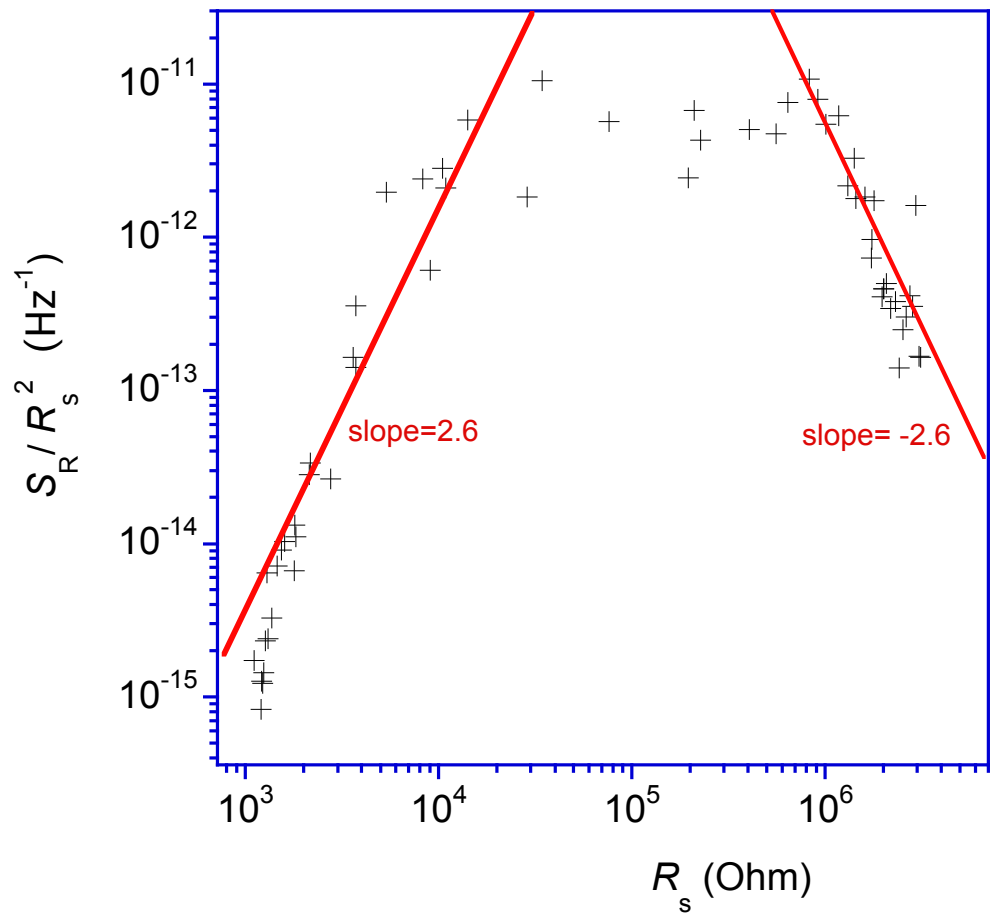

Fig. 5.